\documentclass[twocolumn]{aastex62}
\usepackage{amsxtra}
\usepackage{amsmath}
\usepackage{amssymb}
\usepackage{graphicx}
\usepackage{txfonts}
\usepackage{color}
\usepackage{natbib}

\newcommand{\T}{\ensuremath{\Delta t}}

\newcommand{\ST}{\ensuremath{\Delta s_{\mathrm{\scriptscriptstyle T}}}}
\newcommand{\Ss}{\ensuremath{\Delta s_{\mathrm{\scriptscriptstyle S}}}}
\newcommand{\SM}{\ensuremath{\Delta s_{\mathrm{\scriptscriptstyle M}}}}

\newcommand{\tr}{\ensuremath{\mathbf{b}_n}}

\newcommand{\blos}{\ensuremath{B_{\mathrm{LOS}}\:}}
\newcommand{\W}{\ensuremath{\textbf{W}}}
\renewcommand{\S}{\ensuremath{\mathbf{\Omega}}}

\newcommand{\eref}[1]{equation~(\ref{#1})}
\newcommand{\fref}[1]{Fig.~\ref{#1}}

\defcitealias{paperI}{Paper~I}
\defcitealias{paperII}{Paper~II}

\begin{document}

\title{Stochastic entropy production in the quite Sun magnetic fields}
\email{gorobets@leibniz-kis.de}

\author{A.~Y.~Gorobets}
 
\affil{Kiepenheuer-Institut f\"ur Sonnenphysik, Sch\"oneckstr. 6, D-79104 Freiburg, Germany}
\author{S.~V.~Berdyugina,}
\affil{Kiepenheuer-Institut f\"ur Sonnenphysik, Sch\"oneckstr. 6, D-79104 Freiburg, Germany}

\begin{abstract}
The second law of thermodynamics imposes an increase of macroscopic entropy with time in an isolated system. Microscopically, however, the entropy production can be negative for a single, microscopic realization of a thermodynamic process. The so-called fluctuation theorems provide exact relations between the stochastic entropy consumption and generation. Here, we analyse pixel-to-pixel fluctuations in time of small-scale magnetic fields (SSMF) in the quiet Sun observed with the SDO/HMI instrument. We demonstrate that entropy generated by SSMF obeys the fluctuation theorems. In particular, the SSMF entropy consumption probability is \emph{exactly} exponentially smaller than the SSMF entropy generation probability. This may have fundamental implications for the magnetic energy budget of the Sun.
\end{abstract}


\section{Introduction}\label{introductuon}

Studying spatial and temporal dynamics of small-scale magnetic fields (SSMF) on the solar surface is important for system-wide understanding of the solar magnetism and its role in heating the solar outer atmosphere. A widely used approach to study evolution of SSMF is based on feature tracking. As summarized by \citet{DeForestI} and \citet{DeForestIV}, this approach has significant limitations and subjective biases because of (1) a finite spatial and temporal resolution of observations, and (2) ambiguous boundaries and interactions of SSMF, which are not discrete objects.

In this Letter, we employ a novel formalism which abandons the idea of 'magnetic features' and treats pixel-wise the photospheric magnetic flux density concentrations as homogeneous patches of a flowing substance (field). This approach was recently developed by \citet{paperI} and \citet{paperII} (hereafter \citetalias{paperI} and \citetalias{paperII}), who established that (1) line-of-site (\blos) and longitudinal components of SSMF density evolve as Markov chains \citepalias{paperI}, (2) SSMF is a phenomenological thermodynamic system in a non-equilibrium state (NS, \citetalias{paperII}), and (3) the observed NS thermalizes into a state with a maximum of the information-theoretic entropy \citepalias{paperII}. Here, we develop further this thermodynamic approach to study dynamics of SSMF.

Conceptually, systems in NS interact with their sustained environment (medium, heat bath, or reservoir) by means of influx and efflux of heat, energy, and/or work. If fluxes are time-independent, the system reaches a non-equilibrium steady state (NESS). The environment remains in equilibrium.  This implies that interactions  do not change the macrostate of the environment, but the environment steadily drives the system away from the equilibrium. Together, they form a larger, isolated \emph{total system} \citep[e.g.][]{VandenBroeck20156}.

According to the second law of thermodynamics, entropy of the isolated total system, as a macroscopic quantity, should increase in time. However, it fluctuates due to stochasticity of the microstates, i.e. entropy does not grow monotonically. The probability of the microscopic entropy consumption versus entropy production  is quantified  by the so-called fluctuation theorems (FTs) \cite[e.g.][]{doi:10.1063/1.2012462}.  The FTs and their implications form the framework of stochastic thermodynamics, which is a subject of intensive theoretical and experimental research \citep[e.g. see reviews][]{Harris,Marconi2008111,Seifert_review}.

The goal of this Letter is to deepen our thermodynamic interpretation of SSMF dynamics. First, we apply the NESS model to the observed fluctuating \blos. Then, we show that detailed FT and integral FT are valid on the 'microscopic level'  of discrete time-ordered sequences (trajectories) of \blos at fixed positions of magnetogram images.

\section{Method}
\subsection{Microscopic states and trajectories}

As in  \citetalias{paperII}, we employ the language of thermodynamics to analyse and interpret evolution of photospheric SSMF in a novel way. We read a single solar magnetogram image as a snapshot of microscopic states pixel-wise constituting \blos as an extended macroscopic system. Then, at every pixel, we trace in time its local microscopic dynamics. This is done in the manner of continuous medium sampling by independent detectors (pixels) within the fixed field of view (FOV) \citepalias{paperI}. In time, from image to image, each pixel registers a random sequence of \blos signal values (microstates) $b$  interrupted by noise $\eta$
\begin{align}
\stackrel{\text{\scriptsize{0}}}{|}\xrightarrow[\text{\tiny{TIME}}]{\eta_{0}\eta_{1}\cdots\eta_{0}\eta_{1}\eta_{2}\overbrace{b_0b_1b_2b_3b_4}^\text{\tiny data trajectory}\eta_{0}\eta_{1}\overbrace{b_0b_1\cdots b_{9}}^\text{\tiny data trajectory}\eta_{0}\eta_{1}\cdots\eta_{0}}\stackrel{\text{\scriptsize{T}}}{|}.
\end{align}\noindent
Thus, in our approach, \blos is represented by an ensemble $b(t)$ of finite-length realizations, cf. trajectories in  phase space. The phase space is a finite set of all observed microscopic states. The data trajectories consist of sequential non-noise pixel values within the same FOV during a given observation time $T$. We denote a single trajectory of length $n$ as a time-ordered vector
\begin{align}\label{eq:train}
\tr  &:=\{b_t\}_{t\in[0,n-1]}= \\ \nonumber
&=\bigl\{b_{0}(t_0)\rightarrow b_{1}(t_0+\T) \cdots  \rightarrow b_{n-1}(t_0+(n-1)\T)\bigr\},
\end{align}\noindent
where $t$ is the trajectory's local time index starting at local origin $t_0,\: \T$ is data cadence with $0\le t_0 \le T-\T$, and  $t_{n-1}=t_0+(n-1)\T\le T$.

At a single pixel, the number of trajectories is arbitrary. It depends on the observation time  and the particular solar magnetic field topology. Statistical properties of trajectories are assumed to be homogeneous in space (at least, in the quiet Sun). Hence, trajectories from different pixels contribute to overall statistics equally, i.e. independently on pixel coordinates. Therefore, in our approach, \blos is just  fluctuating in time quantity, with minimal spatial sampling equal to the pixel size and temporal sampling equal to $\T$.

From a thermodynamic perspective, \blos can be characterized by macroscopic values which are quantities averaged over an ensemble of microstates. For instance, its non-equilibrium ensemble entropy $S(t)$ can be written as a microscopic entropy averaged along all trajectories formed by an ensemble of microstates in time
\begin{equation}\label{eq:Entropy_Macro}
S(t)=-\sum_{b\in\S} p(b,t)\ln p(b,t)=-\bigr\langle \ln p\bigr(\tr ,t\bigl)\bigl\rangle
=\bigr\langle s_\mathrm{\scriptscriptstyle S}(t)\bigl\rangle.
\end{equation}\noindent
Here, \begin{equation}\label{eq:Entropy_Micro} s_\mathrm{\scriptscriptstyle S}(t)=s_\mathrm{\scriptscriptstyle S}(\tr ,t):=-\ln p(\tr ,t)\end{equation} is a microscopic system entropy along the trajectory $\tr$, \S{} is a discrete phase space, and $p(\cdot,t)$ is a time-dependent (non-equilibrium) occurrence probability density function (PDF). The microstate PDF is time dependent (see Section~\ref{sec:trajectories}), and trajectories are considered as being distributed stationary $p(\tr,t)\equiv p(\tr)$.
The microscopic extension of the entropy \eref{eq:Entropy_Micro} was justified by \citet{Crooks}, \citet{PhysRevE.65.016102}, and \citet{Seifert_trajectories}. The non-equilibrium of \blos was examined in terms of transition probability matrix asymmetry (detailed balance violation) in \citetalias{paperII}.  At the same time, trajectories obey the Markov property  \citepalias{paperI}, which allows us to estimate $p(\tr)$. Thus, the joint occurrence PDF for $\tr $ (with $n$ time-ordered independent arguments) can be explicitly computed by the following factorization:
\begin{align}\label{eq:factor}
p(\tr ):=&p_n(b_{0}, b_{1},b_{2} \cdots b_{n-1})\\
=&p(b_0,t_0)w(b_{1}|b_{0})w(b_{2}|b_{1})\cdots w(b_{n-1} | b_{n-2})\nonumber.
\end{align}\noindent
Here, $w(b_j|b_i)$ is the $\T$-step conditional PDF of a random transition from the microstate $b_i$ into $b_j$, and $p(b_0,t_0)$ is the occurrence PDF of $b_0$ at the trajectory's initial time $t_0$. The state space  $\S$ (all possible values of binned \blos) has a finite number $\Omega$ of elements determined by the bin size. The bin size is computed as in \citetalias{paperII}. The PDF $p(b,t)$ and the $\Omega\times \Omega$ transition matrix $\W$ of $\T$-step transition probabilities with the elements $w_{ij}=w(b_j|b_i)\, i,j\in[1,\Omega]$ provide complete information on the statistics of the ensemble $b(t)$ due to the Markov property \eref{eq:factor}. In this way, pixel by pixel, we collect all necessary information for the microscopic description of the non-equilibrium \blos system.

\subsection{Entropy production along trajectories}

Trajectories allow us to characterize \blos as a thermodynamic system, which is in contact with its environment (medium), using a microscopic entropy flux $\Delta s$. Here, we define such a flux for the system ($\Ss$) and its medium ($\SM$), as well as their total ($\ST$).

The system's entropy (equation \ref{eq:Entropy_Micro}) variation is simply a change of microscopic entropy between the initial and final states of the trajectory
\begin{flalign}\label{eq:ss} \nonumber
\Ss(\tr ) :=s_\mathrm{\scriptscriptstyle S}(t_{n-1})-s_\mathrm{\scriptscriptstyle S}(t_0)
                &=\ln p(b_0,t_0)-\ln p(b_{n-1},t_{n-1})\\
                &=\ln\left[\frac{p\left(b_0,t_0\right)}{p\left(b_{n-1},t_{n-1}\right)}\right].
\end{flalign}

The medium's entropy variation is defined by a single microscopic $\T$-transition $b_i\rightarrow b_j$
\begin{flalign}\label{eq:sm}
\SM := \ln \left[\frac{w(b_{j}|b_i)}{w(b_i|b_{j})}\right].
\end{flalign}\noindent
This is a relative measure of the forward transition with respect to its reversal. The latter occurs when the environment statistically influences the system so that it returns to the previous state.

Finally, the \textit{total entropy production} of the system together with its medium $\ST=\Ss+\SM$  is defined for a Markovian $\tr $ as follows:
\begin{flalign}\label{eq:str}
\ST(\tr )=&\ln \left[\frac{p(b_0, t_0)}{p(b_{n-1},t_{n-1})} \prod\limits_{k=0}^{n-2}\frac{w(b_{k+1}|b_{k})}
{w(b_{k}|b_{k+1})}\right]\\ \label{eq:str2}
=&\ln\left[\frac{p_n\left(b_0,b_1,b_2\cdots b_{n-1}\right)}{p_n\left(b_{n-1}\cdots b_{2},b_1,b_0\right)}\right]=\ln\left[ \frac{p(\tr )}{p(\tilde{\mathbf{b}}_n)}\right],
\end{flalign}\noindent
where $\tilde{\mathbf{b}}_n$ is the time-reversed trajectory of $\tr $, i.e. $\tr $ is read backwards. The $p(\tr )$ and $p(\tilde{\mathbf{b}}_n)$ differ by the order of the arguments in correspondent $w(\cdot|\cdot)$ of \eref{eq:str} and by PDFs of the corresponding initial states at $t_0$ and $t_{n-1}$. How to compute PDFs of the initial states is explained in Section~\ref{sec:trajectories}.

We note that the environment interacts with the system by means of some kind of energy ('heat'), which is a path function rather than state variable. Thus, the environment influences the probabilistic evolution between states $w(b_j|b_i)$ but not the states $b_{i(j)}$ themselves \citep[e.g.][]{Hinrichsen}. In the context of \blos, we define its environment as all physical processes and conditions affecting \blos by means of energy transfer and/or conversion, which are not considered and/or inaccessible by our measurements of \blos\!.

\subsection{Fluctuation theorems}

The microscopic entropy production $\ST$ obeys well-defined statistical relations (e.g.  \citealt[]{Klages2013}; \citealt{VandenBroeck20156}), which we verify in our analysis.

\subsubsection{Detailed fluctuation theorem (DFT)}

DFT defines the exact relation between probabilities for the positive and negative entropy production of the same magnitude  $|\ST|$:
\begin{equation}
 \frac{p(\ST>0)}{p(\ST<0)} = e^{\ST>0}.\label{eq:FT}
\end{equation}
This exponential dominance of entropy-generating trajectories over entropy-consuming trajectories quantifies the microscopic \emph{irreversibility} in stochastic dynamics of the system.
With the normalisation condition for $p(\tr )$, DFT leads to equation for the ensemble average (over all trajectories)
\begin{equation}
 \langle e^{-\ST}\rangle=1. \label{eq:iFT2}
\end{equation}\noindent
Using the Jensen inequality $e^{\langle x \rangle} \le \langle e^x\rangle$, we obtain the second law of thermodynamics: $\langle\ST\rangle\ge0$, which is the macroscopic consequence of the exponential smallness of the entropy consumption constrained by DFT.
\subsubsection{Integral fluctuation theorem (IFT)}

A set of trajectories of equal $n$ can be split into two disjoint subsets with respect to the sign of $\ST(\tr )$. IFT predicts the probability ratio of entropy-consuming trajectories ($\ST(\tr )<0$) to entropy-generating trajectories ($\ST(\tr )>0$):
\begin{equation}
 \frac{p(\tr |\ST(\tr )<0)}{p(\tr |\ST(\tr ) >0)} = \left\langle e^{-(\ST(\tr ) >0)}\right\rangle
 .\label{eq:iFT}
\end{equation}\noindent
The trajectory length $n$ in this equation is a parameter, i.e. the right-hand side averaging is carried over a subset of entropy-generating trajectories of particular $n$ \citep[e.g.][]{2004PhRvL..92n0601C}.

\section{Observational data}\label{section:data}

We analyse uninterrupted SDO/HMI observations of the quiet Sun \blos near the disc centre \citep{2012SoPh..275..207S,2012SoPh..275..229S}. We have chosen a sequence of $15335$ magnetograms in the Fe I 6173 \AA{} line from 2017 February 2, 00:00:23~UT until 2017 February 8, 23:58:53~UT with a cadence of $\T=45~\mathrm{s}$ and angular resolution of $1\arcsec$. The images were preprocessed with the \texttt{hmi\_prep.pro} procedure from the SolarSoft package, and then cropped to the $400\times400$ pixel area around the disc centre. Here, we focus on \blos\!\!, since the available data have sufficient temporal resolution only for the \blos component of the vector magnetic field in SSMF. However, the fundamentally important Markov property for transverse component was also demonstrated for high resolution measurements of SSMF \citepalias{paperI}.

To minimize a possibly induced 'memory effect' by the data pipeline preprocessing, we use \emph{near-real-time} series \textit{hmi.M\_45s\_nrt}. These data are linearly interpolated filtergrams over three temporal intervals, in contrast to the $sinc$-function spatio-temporal interpolation over five intervals in regular (non-nrt) data series \citep{2011SoPh..269..269M, 2017ApJ...834...26K}. The slightly overestimated noise level of $\sigma=10.3~\mathrm{Mx~cm^{-2}}$ \citep{SDO-tech} for the disc centre is used, as in \citetalias{paperII}.

It is known that the orbital velocity of the SDO satellite relative to the Sun is a periodic function with the amplitude of $\pm3~\mathrm{km~s^{-1}}$ (FITS header keyword \texttt{OBS\_VR}). \textsc{HMI} line-of-sight observables are hence affected by periodic Doppler shifts of the spectral line due to the satellite's orbit. Together with the solar rotation, this satellite's geosynchronous orbital motion leads to temporal and spatial inhomogeneous instrumental variations of inferred magnetic fields over the solar disc with the period of 24 h \citep[][]{Hoeksema}.

To have statistically homogeneous data, we select images with a uniform noise pattern within the cropped area. The selection is made by estimating $\Delta \eta(t)=\langle| \eta(t) - \eta(t+\T)|\rangle_{xy}$, where $\eta(t)$ and $\eta(t+\T)$ are noise pixels ($|\blos|<2\sigma$) with the same coordinates $\{xy\}$ in two subsequent images, while averaging is spatial. We have found that discrepancy in $\Delta \eta(t)$ computed separately for the west and east disc parts of the cropped area is minimal when the satellite's orbital velocity ranges  $0.15-1.6~\mathrm{km~s^{-1}}$. This range of velocities defines $\approx 2~\mathrm{h}$- intervals in the magnetogram data for each cycle of the SDO satellite turnover. In total, we have $16$ such intervals during the selected observational sequence.

\section{Data Analysis and Results}

\subsection{Time dependence of the microstate PDF}\label{sec:trajectories}

When assembling trajectories from magnetogram time series, we seek for pixels with persistent non-noise signals.  Our algorithm is identical to the one introduced in \citetalias{paperI} and \citetalias{paperII}.

First, we independently choose pixels above the noise cut-off threshold in two images at $t$ and $t+\T$. Then, in the same two images, we select pixel pairs which are above the cut-off simultaneously -- they become contributors to trajectories at corresponding pixel positions. This iteration is repeated after shifting the two-image-stencil to the next position in time by the $\T$-step, so that the next pair of images is at times $t+\T$ and $t+2\T$. We take care to avoid selecting samples twice, i.e. two iterations with one stencil shift described above give three (not four) samples in the trajectory, because the same image appears twice in the stencil per two iterations. Finally, to make samples $b_i(t)$ in $b_j(t+\T)$ being distributed with the same mean, the local three-image average is subtracted from the current image pair contributing to the trajectory.

While gathering statistics, by sequential shift of the stencil through the whole sequence of magnetograms,
we found a non-vanishing discrepancy $\partial_t p=\left[p(b,t+\T)-p(b,t)\right]/\T\neq0$, known in the literature as the probability current. By construction, it is assumed to be stationary for $t\in[0,T]$, and it corresponds to the macroscopic flux (or fluxes) constituting the NESS macrostate. Physically, it is related to steady processes altering \blos on $\T$ time-scale in our data.

According to our algorithm, samples in trajectories belong to alternating NESSs characterized by $p(b,t)$ and $p(b,t+\T)$. Therefore, in \eref{eq:str} $p(b_{n-1},t_{n-1})=p(b,t+\T)$ which is the $b$ occurrence PDF in the $\T$-stable non-noise pixels in every second magnetogram of the sequence, and $p(b_0,t_0)=p(b,t)$.

\subsection{Test for DFT}

We verify and compare DFT for three sets of trajectories $\tr $: (1) simulated with normally distributed $b_t$ and $n$, (2) assembled from HMI noise pixels (signal $\le2\sigma$), and (3) assembled from HMI data with the signal $\ge4\sigma$. In the last two sets, $n$ is defined by the noise--signal intermittency.

\begin{figure}\centering \includegraphics[scale=0.8]{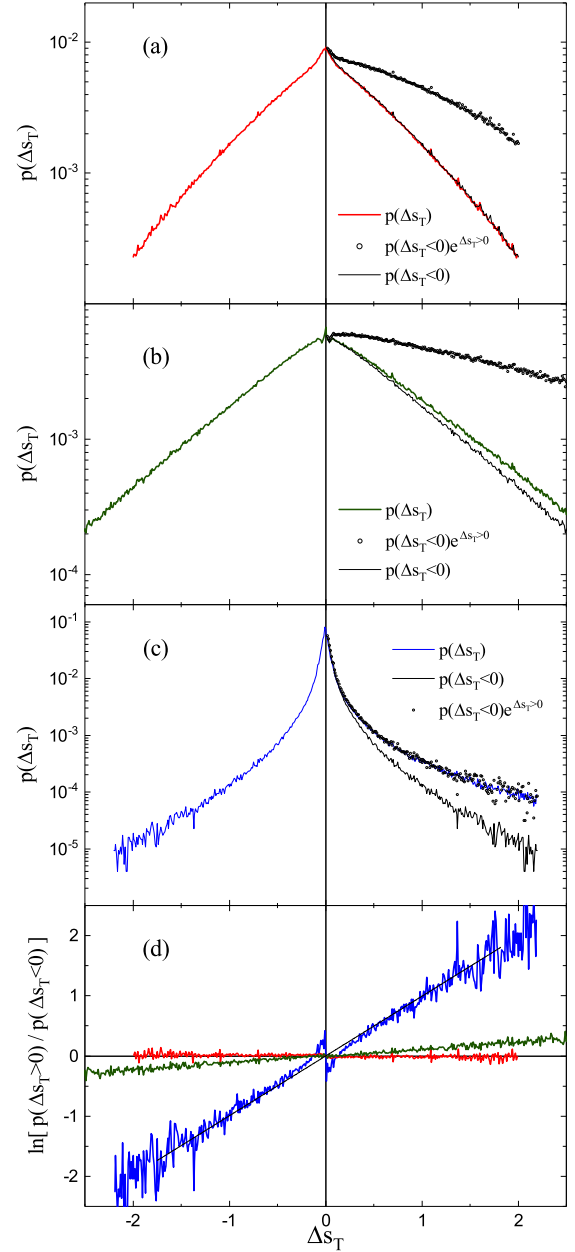}
\caption{Empirical tests of DFT: $p(\ST)$ (solid colour lines) is plotted together with its mirrored half-plot for $\ST<0$ (black line), which is then multiplied by $\exp({\ST>0})$  according to \eref{eq:FT} and shown by circles. {Panel A}: DFT for simulated Gaussian trajectories. {Panel B}: DFT for observed noise trajectories. {Panel C}: DFT for observed \blos trajectories with the signal above the $4\sigma$ noise cut-off. {Panel D}: Logarithm of the left side of \eref{eq:FT} versus $\ST$ (same notation as in the upper panels). The orthogonal distance regression fit $y=\alpha+\beta x$ (black) applied to the non-noise data with $\alpha=-0.00065\pm0.00918$ and $\beta=0.99287\pm0.00891$ strongly supports DFT.}
\label{fig1}
\end{figure}

These sets define the time-forward structure of the corresponding $\W$, but none of them provide reverse sequences $\tilde{\mathbf{b}}_n$ explicitly. Therefore, time reversals in equations~(\ref{eq:str}) and (\ref{eq:str2}) are computed formally. It is remarkable that natural time-forward stochastic dynamics of microstates in sets (2) and (3) reveals such time-forward trajectories whose reversals lead to existence of microscopic $\ST<0$, as shown in \fref{fig1}.

The set (1) of Gaussian trajectories with statistically independent microstate occurrences has a symmetric entropy production PDF $p(\ST)$, as expected (see \fref{fig1}A).

The observed noise in the set (2) should in principle have a Gaussian nature of statistical independence. However, it appears that residual biases in data processing and instrumental effects break the time reversal symmetry. The $p(\ST)$ for HMI noise trajectories shown in \fref{fig1}B reveal a noticeable discrepancy between $p(\ST>0)$ and $p(\ST<0)$ (black line versus green line).

The \blos signal trajectories in the set (3) validate DFT, as demonstrated in \fref{fig1}C. The symbols depicting $p(\ST<0)\exp(\ST>0)$) remarkably follow the solid curve of $p(\ST>0)$. This confirms the relation given by \eref{eq:FT}. At the origin, $p(\ST)$ is symmetric, implying that the irreversibility is detectable for relatively large $\ST$ corresponding to quite prolonged times (see discussion of \fref{fig2}B).

In \fref{fig1}D, we plot the logarithm of \eref{eq:FT}. It should be a linear function of $\ST$ with the slope of $1$ for larger $\ST$, where asymmetry of $p(\ST)$ is the strongest. The fit (black line) confirms the diagonal slope of the logarithm of \eref{eq:FT} for the \blos signal trajectories. This behaviour appears to be robust against variations in the number of $2\,\mathrm{h}$-intervals used in the analysis as well as against the noise cut-off level.

The green line in \fref{fig1}B and \fref{fig1}D corresponds to the noise trajectories whose $p(b,t)$ and $\W$ were estimated during only one $2\,\mathrm{h}$-interval. The slope of the green line in \fref{fig1}D is $0.104\pm0.001$. When increasing the number of $2\,\mathrm{h}$-intervals for the noise trajectories, the slope increases by $\approx0.066$ per interval, up to the limiting value of $1$. A possible source of such a data-like behaviour in the noise may be hidden in HMI data calibration techniques, such as discussed by \citet{Hoeksema}. Alternatively, the noise may contain some weak traces of the \blos signal. If the latter is true, we may gradually amplify the statistical significance of this weak signal by increasing the number of $2\,\mathrm{h}$-intervals.

\subsection{Test for IFT}

\begin{figure}\centering \includegraphics[scale=.6]{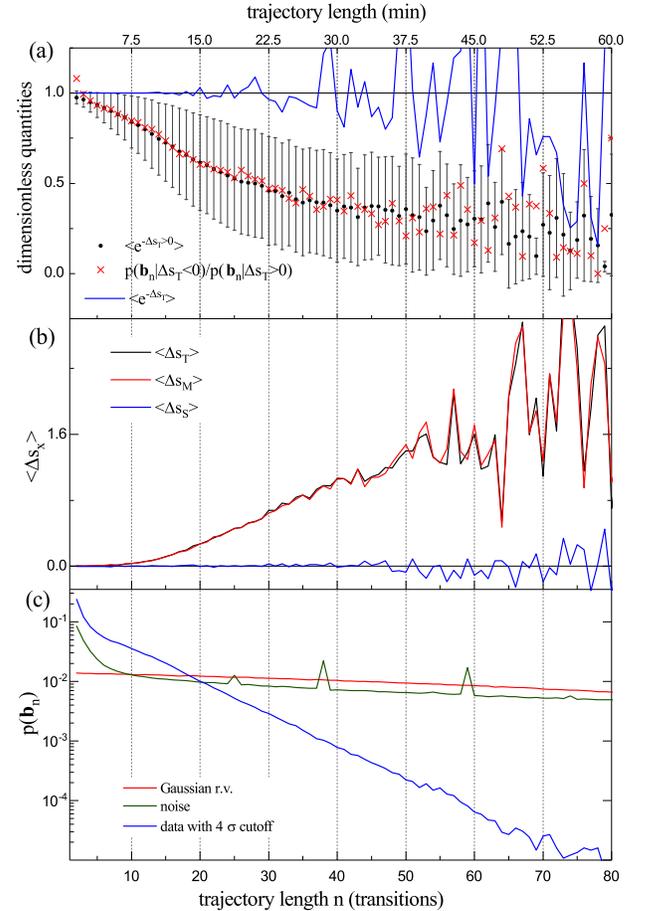}
\caption{Empirical test of IFT. {Panel A}: The left-hand side of \eref{eq:iFT} (circles), its right-hand side (crosses), and \eref{eq:iFT2} (blue line) versus the trajectory length $n$ (duration in minutes is shown at the top axis). {Panel B}: Average entropy productions $\langle\ST\rangle$ (black), $\langle\Ss\rangle$ (blue), and $\langle\SM\rangle$ (red) as functions of $n$. The medium entropy production is computed as the second summand in \eref{eq:str}:~$\SM(\tr )=\ln\prod_{k=0}^{n-2}\left\{w(b_{k+1}|b_{k}) /w(b_{k}|b_{k+1})\right\}$. {Panel C}: The occurrence PDF for Gaussian trajectories (red), noise trajectories (green), and \blos signal (blue). See the text for more details.}\label{fig2}
\end{figure}

To validate IFT, we compute the left-hand and right-hand sides of \eref{eq:iFT} as independent functions for trajectories of a particular $n$. In \fref{fig2}A, the result for \blos data trajectories is shown as red crosses (left-hand side) and black circles (right-hand side). They coincide well within errors. The errors are computed as standard deviations of $\ST$ for entropy-generating trajectories. The blue line in \fref{fig2}A corresponds to \eref{eq:iFT2}. On average, it is close to 1, especially for shorter trajectories, which occur more frequently.

The exponential excess of entropy-generating trajectories must lead to a net increase of entropy with time in the total system in accord with the second law of thermodynamics. To verify this, we plot time-averages of $\ST$, $\Ss$ and $\SM$ as functions of $n$ in \fref{fig2}B, which are the averages over trajectories of time length $n\T$. First, the plot demonstrates that $\Ss$ averages to zero (blue line). This is because the system in NESS exchanges energy with the medium at a constant rate, i.e. $dS/dt=0$.  And second, non-negative increasing function $\langle\SM\rangle$ (red line) reveals imbalanced (non-equilibrium) transitions in the system \citep[see][section 2]{Hinrichsen}, and, at the same time, increase of entropy in the environment, which is the energetic cost to keep the dissipating system in the NESS. According to the NESS model, the medium stays in equilibrium, and hence it should balance own reaction to changes in the system on time-scales much shorter than required to change the macroscopic state of the medium. This time-scale separation is the underlying mechanism for the existence of NESS \cite[e.g. see][section 1.2]{Seifert_review}.

Since $\langle\SM\rangle$ grows and $\langle\Ss\rangle$ fluctuates near zero, $\langle\ST\rangle=\langle\Ss\rangle+\langle\SM\rangle$ (black line) grows almost the same as $\langle\SM\rangle$. However, fluctuations of $\Ss$ are crucial on the microscopic level, since the DFT is valid only for $\ST$ as verified by our analysis.

The macroscopic fluctuating  entropy $S(t)$ as the average of microscopic entropies along trajectories requires a certain observation time, over which the time-average manifests its growth, say a mesoscopic time-scale. Larger system, shorter is its mesoscopic time-scale. In our analysis, the existence of the mesoscale is implicitly shown by the initially slow increase of $\langle\ST\rangle$ in \fref{fig2}B. For small $n$ around the peak of $p(\tr )$, $\langle\ST\rangle$ is indeed negligibly small.

We note that estimates in \fref{fig2}A and \fref{fig2}B appear to diverge with time. This is because longer data trajectories are less frequent, as shown in \fref{fig2}C.

\section{Summary and Conclusions}\label{discussion}

We have shown that the fundamental FTs discovered both theoretically and experimentally in the framework of stochastic thermodynamics are also valid for evolution of solar SSMF. In particular, \blos fluctuations in SDO/HMI quiet photosphere data with the pixel size of 0.5 arcsec and cadence of $45~\mathrm{s}$ firmly satisfy both DFT and IFT. We have also found a non-zero probability current as a systematic statistical difference in PDFs of the same spatial structure of \blos. Such probability currents are used to identify NESSs \citep[e.g.][]{Zia}.

Thus, evolution of SSMF in the solar photosphere can be considered as a stochastic NESS ensemble governed by FT relationships on the microscopic level and by the second law of thermodynamics on the macroscopic one. In the essence, our approach is based on the information-like definition of the entropy, i.e. it is dimensionless  \citep{6773024, Cover:2006:EIT:1146355}.  Whether and how the studied here entropy production by SSMF may be related to heating of the upper solar atmosphere and general energy transfer is to be clarified. We suggest that a similar analysis is carried out for other small-scale phenomena in the solar chromosphere, transition region, and corona. In addition, such an analysis can be applied to MHD solar simulations to verify whether simulated SSMF have the same statistical properties as observed ones.

Our DFT test has also revealed that noise in HMI data changes its statistical properties on the time-scale longer than one SDO orbit. For relatively short-time intervals (few hours), noise samples behave as being statistically independent.

This work advances our new approach to the analysis of evolution of solar magnetic fields based on phenomenological thermodynamics and information-theoretic definition of entropy. We believe that it has a potential for applications to other astrophysical complex phenomena observed in continuous media (fields), as well as for revealing systematic errors in data.

\section{Acknowledgements}

This work was supported by the European Research Council Advanced Grant HotMol (ERC-2011-AdG 291659).
\textsc{SDO} is a mission for NASA's Living With a Star (LWS) program. The \textsc{SDO/HMI} data were provided by the Joint Science Operation Center (JSOC).


\end{document}